\documentstyle[12pt,epsfig]{article}
\title{Refined Spectral Method as an extremely accurate technique for solving 2D time-independent Schr\"{o}dinger equation}
\author{P. Pedram\thanks{Email: pedram@sbu.ac.ir}, M. Mirzaei, and S. S. Gousheh
\\ {\small Department of Physics, Shahid Beheshti University,
Evin, Tehran 19839, Iran}}
\begin{document}
\maketitle \baselineskip 24pt
\begin{abstract}
We present a refinement of the Spectral Method by incorporating an
optimization method into it and generalize it to two space
dimensions. We then apply this Refined Spectral Method as an
extremely accurate technique for finding the bound states of the
two dimensional time-independent Schr\"{o}dinger equation. We
first illustrate the use of this method on an exactly solvable
case and then use it on a case which is not so. This method is
very simple to program, fast, extremely accurate ({\it e.g.} a
relative error of $10^{-15}$ is easily obtainable in two
dimensions), very robust and stable. Most importantly, one can
obtain the energies and the wave functions of as many of the bound
states as desired with a single run of the algorithm.

\vspace{5mm} {\it PACS numbers:} 02.70.Hm, 03.65.Ge \vspace{15mm}
\end{abstract}
\maketitle
\section{Introduction}
Eighty years after the birth of quantum mechanics
\cite{Schrodinger}, the Schr\"{o}dinger's famous equation still
remains a subject for numerous studies, aiming at extending its
field of applications and at developing more efficient analytic
and approximation methods for obtaining its solutions. There has
always been a remarkable interest in studying exactly solvable
Schr\"{o}dinger equations which has been found for only a very
limited number of potentials, most of them being classified
already by Infeld and Hull \cite{Infeld} on the basis of the
Schr\"{o}dinger factorization method \cite{factorization}, which
in turn appeared to be a rediscovery of the formalism stated
nearly 120 years ago by Darboux \cite{Darboux}. However, a vast
majority of the problems of physical interest do not fall in the
above category when we formulate a more or less realistic model
for them. Then we have to resort to approximation techniques which
can be analytic or numeric. The Schr\"{o}dinger equation can
always be solved numerically, which nowadays seems elementary, in
view of the immensely increased computational power. However, even
in this simplest case, the success of applying any direct
numerical integration method depends on the quality of initial
guesses for the boundary conditions and energy eigenvalues.
Moreover, one usually encounters difficulties with the intrinsic
instabilities of typical problems, and rarely with the existence
of actual solutions which posses rapid oscillation. The need for
evermore accurate and efficient numerical methods for solving
problems of physical interest have stimulated development of more
sophisticated integration approaches, {\it e.g.} embedded
exponentially-fitted Runge-Kutta \cite{Runge-Kutta} and
dissipative Numerov-type \cite{Numerov} methods, as well as
interesting techniques, such as a relaxation approach \cite{Relax}
based on the Henyey algorithm \cite{Henyey}, an adaptive basis set
using a hierarchical finite element method \cite{hierarchical},
and an approach based on microgenetic algorithm
\cite{microgenetic}, which is a variation of a global optimization
strategy proposed by Holland \cite{Holland}. Most of these methods
are either completely designed for the one-dimensional cases or
optimized so. Few general methods are readily available for higher
dimensional cases, \textit{e.g} Finite Element Method (FEM),
Finite Difference Method (FDM), Relaxation Method, Spectral
Element Method.

Here we extend the Refined Spectral Method (RSM), which is
introduced in Ref. \cite{RSM}, as a numerical method to solve the
higher dimensional schr\"{o}dinger equations. There, we first
refined the Spectral Method (SM) \cite{SP1} for one-dimensional
cases by incorporating an optimization procedure into it, and then
tested the results obtained by our method against the
corresponding values of an exactly solvable case. We showed that
this method can be extremely accurate, ({\it e.g.} errors of order
$10^{-130}$), and has the following advantages: It is very simple,
fast, very robust and stable, {\it i.e.} it does not have the
instability problems due to the usual existence of divergent
solutions of most physical problems. These problems usually
produce difficulties for the spatial integration routines such as
FDM and FEM. Finally, and perhaps most importantly, we can obtain
the wave functions and energies of as many of the bound states as
desired with a single run of the algorithm. Spectral Method,
consists of first choosing a complete orthonormal set of
eigenstates of a, preferably relevant, hermitian operator to be
used as a suitable basis for our solution. For this numerical
method we obviously can not choose the whole set of the complete
basis, as these are usually infinite. Therefore we make the
approximation of representing the solution by a superposition of
only a finite number of the basis functions. By substituting this
approximate solution into the differential equation, a matrix
equation is obtained. The energies and expansion coefficients of
these approximate solutions could be determined by the eigenvalues
and eigenfunctions of this matrix, respectively. In the Spectral
Method the concentration is on the basis functions and we expect
the final numerical solution to be approximately independent of
the actual basis used. Moreover in this method, the refinement of
the solution is accomplished by choosing a larger set of basis
functions, rather than choosing more grid points, as in the
numerical integration methods. For more detailed explanation on
this subject, in particular different branches of SM, including
the commonly used Pseudo-Spectral Method, and its historical
development see for example Ref. \cite{SP1}. For an interesting
application of this method to the double well potential see for
example Ref. \cite{RSM2}.

The remainder of this paper is organized as follows. In Section 2,
we present the underlying theoretical bases for the formulation of
the RSM and introduce our optimization procedure in
two-dimensions. In Section 3, we first use this method for the 2D
Simple Harmonic Oscillator (2D-SHO), which is an exactly solvable
problem, to illustrate  and test the method. In Section 4, we
apply this method to an interesting 2D problem which could be
relevant to QCD and is not exactly solvable. In Section 5, we
state our conclusions.

\section{The Refined Spectral Method}
Let us consider the 2-D time-independent Schr\"{o}dinger equation,
\begin{equation}\label{Schrodinger}
-\frac{\hbar^2}{2m}\left(\frac{d^2\psi(x,y)}{dx^2}+\frac{d^2\psi(x,y)}{dy^2}\right)+U(x,y)\psi(x,y)=E\psi(x,y),
\end{equation}
where $m$, $U(x,y)$, and $E$ stand for the reduced mass, potential
energy, and energy, respectively. Obviously, This is an eigenvalue
problem with eigenfunction $\psi(x,y)$ and eigenvalue $E$.
Throughout this paper, we only examine the bound states of this
problem, i.e. the states which are the square integrable.
Therefore the general eigenvalue problem that we want to solve can
be cast in the form of a linear elliptic PDE one that can be
written as,
\begin{equation}\label{ODE}
-\left(\frac{d^2\psi(x,y)}{dx^2}+\frac{d^2\psi(x,y)}{dy^2}\right)+\hat{f}(x,y)\psi(x,y)=\varepsilon\,\psi(x,y),
\end{equation}
where,
\begin{eqnarray}
\hat f(x,y)=\frac{2m}{\hbar^2}\, U(x,y),\label{f}
\hspace{2cm}\varepsilon=\frac{2m}{\hbar^2}\, E.
\end{eqnarray}
The configuration space for most physical problems are defined by
$-\infty<x,y<\infty$. We make the approximation of constraining
the domain to $-L_x/2<x<L_x/2$ and $-L_y/2<y<L_y/2$. As shall be
explained later, first of all, this constraining of the domain is
absolutely crucial for our method, and secondly does not
necessarily pose a loss of accuracy: The use of a finite domain is
necessary since we need to choose a finite subspace of a countably
infinite basis. Moreover, since the bound states have compact
support, a finite region suffices, and the choices of $L_x$ and
$L_y$ are in fact the essential part of our optimization
procedure. As mentioned before, any complete orthonormal set can
be used for the SM. We use the Fourier series basis as an example.
For this particular basis, we find it convenient to shift the
domain to $0<x<L_x$ and $0<y<L_y$. In particular, we need to shift
the potential energy functions also. This means that we can expand
the solution as,
\begin{eqnarray} \psi(x,y)=\sum_{m,n=1}^{\infty} A_{m,n}
\,\,\, \sin\left(\frac{m \pi x}{L_x}\right)\,\sin\left(\frac{n \pi
y}{L_y}\right). \label{eqpsitrigonometric}
\end{eqnarray}
We can also make the following expansion,
\begin{eqnarray}
\hat f(x,y) \psi(x,y)=\sum_{m,n} B_{m,n} \,\,\,\sin\left(\frac{m
\pi x}{L_x}\right)\,\sin\left(\frac{n \pi
y}{L_y}\right),\label{eqV}
\end{eqnarray}
where $B_{m,n}$ are coefficients that can be determined once $\hat
f(x,y)$ is specified. By substituting Eqs.\
(\ref{eqpsitrigonometric},\ref{eqV}) into Eq.\ (\ref{ODE}) and
using the differential equation of the Fourier basis we obtain,
\begin{eqnarray}
\hspace{-2cm}\sum_{m,n}\left[\left(\left(\frac{m
\pi}{L_x}\right)^2+\left(\frac{n
\pi}{L_y}\right)^2-\varepsilon\right)
A_{m,n}+B_{m,n}\right]\sin\left(\frac{m \pi
x}{L_x}\right)\,\sin\left(\frac{n \pi
y}{L_y}\right)=0.\label{eqAB1}
\end{eqnarray}
Because of the linear independence of $\sin\left(\frac{m \pi
x}{L_x}\right)$ and $\sin\left(\frac{n \pi y}{L_y}\right)$, every
term in the summation must satisfy,
\begin{eqnarray}
\left(\left(\frac{m \pi}{L_x}\right)^2+\left(\frac{n
\pi}{L_y}\right)^2\right) A_{m,n}+B_{m,n}=\varepsilon\,
A_{m,n}.\label{eqAB2}
\end{eqnarray}
It only remains to determine the matrix $B$. Using Eq.\
(\ref{eqV}) and Eq.\ (\ref{eqpsitrigonometric}) we have,
\begin{eqnarray}
\sum_{m,n} B_{m,n} \sin\left(\frac{m \pi
x}{L_x}\right)\,\sin\left(\frac{n \pi y}{L_y}\right)=\sum_{m,n}
A_{m,n} \hat f(x,y) \sin\left(\frac{m \pi
x}{L_x}\right)\,\sin\left(\frac{n \pi y}{L_y}\right).
\end{eqnarray}
By multiplying both sides of the above equation by
$\sin\left(\frac{m \pi x}{L_x}\right)\,\sin\left(\frac{n \pi
y}{L_y}\right)$ and integrating over the $x,y$-space and using the
orthonormality condition of the basis functions, one finds,
\begin{eqnarray}
B_{m,n}= \sum_{m',n'} C_{m,m',n,n'}\,\, A_{m',n'},
\end{eqnarray}
where,
\begin{eqnarray}
C_{m,m',n,n'}=(\frac{4}{L_xL_y})\int_{0}^{L_x}\int_{0}^{L_y}
\sin\left(\frac{m \pi x}{L_x}\right)\,\sin\left(\frac{n \pi
y}{L_y}\right)\hat f(x,y)\,\,\, \sin\left(\frac{m' \pi
x}{L_x}\right)\,\sin\left(\frac{n' \pi y}{L_y}\right) dxdy.
\end{eqnarray}
Therefore we can rewrite Eq.\ (\ref{eqAB2}) as,
\begin{eqnarray}
\left(\left(\frac{m \pi}{L_x}\right)^2+\left(\frac{n
\pi}{L_y}\right)^2\right) A_{m,n}+ \sum_{m',n'} C_{m,m',n,n'}\,\,
A_{m',n'}=\varepsilon\, A_{m,n}.\label{eqAC}
\end{eqnarray}
It is obvious that the presence of the operator $\hat f(x,y)$ in
Eq.\ (\ref{ODE}), leads to nonzero coefficients $C_{m,m',n,n'}$ in
Eq.\ (\ref{eqAC}), which in principle could couple all of the
matrix elements of $A$. It is easy to see that the more basis
functions we include, the closer our solution will be to the exact
one. We select a finite subset of the basis functions
\textit{i.e.} the first $N^2$ ones, by letting the index $m$ and
$n$ run from 1 to $N$ in the summations. Then we replace the
square matrix $A$ with a column vector $A'$ with $N^2$ elements,
so that any element of $A$ corresponds to one element of $A'$.
With this replacement, Eq. (\ref{eqAC}) can be written as,
\begin{eqnarray}
D\, A'=\varepsilon\, A', \label{eqmatrix}
\end{eqnarray}
where $D$ is a square matrix with $(N^2) \times (N^2)$ elements.
Its elements can be obtained from Eq.\ (\ref{eqAC}). The
eigenvalues and eigenfunctions of the Schr\"{o}dinger equation are
approximately equal to the corresponding quantities of the matrix
$D$. That is the solution to this matrix equation simultaneously
yields $N^2$ sought after eigenstates and eigenvalues. The only
problem which remains is to solve the eigenvalue problem Eq.\
(\ref{eqmatrix}), and to control the round-off errors. This is
often a serious issue for the usual spatial integration method
using double precision. However, we can easily overcome this
problem and obtain a very high precision. Using RSM in 1D
accuracies of order 100 significant digits are very easily
accomplishable while in 2D 10 significant digits are obtained
using the same computation time. This can be implemented, for
instance with MATHEMATICA, using the instruction
`Set[Precision[...,20]', for example, to set a precision of 20
digits for the numbers. This method, in principle, allows us to
obtain the eigenvalues and eigenvectors with a maximum precision
of 20 digits (using enough basis elements).

Now we can introduce our optimization procedure. We are free to
adjust two parameters: $N$, the number of basis elements used and
the lengths of the spatial region, $L_x$ and $L_y$. These lengths
should be preferably larger than spatial spreading of all the
sought after wave functions. However, if $L_x$ and $L_y$ are
chosen to be too large we loose overall accuracy. After fixing
these lengths, any desired accuracy can be obtained with a
suitable choice of $N$. As we shall show, the error decreases
extremely rapidly as the number of basis elements is increased.
However, it is important to note that for each $N$, $L_x$ and
$L_y$ have to be properly adjusted. We shall denote these optimal
quantities by $\hat{L_x}(N)$ and $\hat{L_y}(N)$. We have come up
with a method to determine these quantities: For a few fixed
values of $N$ we compute $E(N,L_x,L_y)$ which invariably has an
minimum point. Therefore, all we have to do is to compute the
position of these minimum points and compute an interpolating
function for obtaining $\hat{L_x}(N)$ and $\hat{L_y}(N)$.
Obviously the more points we choose the better our results will
be. As we shall see, the addition of this refinement can have
dramatic consequences.

Computation of the relative error in the exactly solvable cases is
straightforward. For example for computing the relative error of
the eigenvalue, denoted by $\delta _E$, we only need to find the
absolute value of the difference between the result and the exact
one and divide by the latter. For cases which are not exactly
solvable, we compute the difference between the eigenvalues for a
given $N$ and those obtained with $N+1$, both lying on the
$\hat{L_x}(N)$ and $\hat{L_y}(N)$ curves. We shall denote the
error computed by this procedure $\hat{\delta}_E$. We have
computed $\hat{L_x}(N)$ and $\hat{L_y}(N)$ for all cases, and
subsequently computed the eigenfunctions, eigenvalues and their
errors using this method, and  checked their validity in the
exactly solvable case of 2D-SHO. Obviously to obtain consistent
results we have to keep the same precision throughout the
calculations.

\section{2D Simple Harmonic Oscillator}
In this section, for illustrative purposes, we apply RSM to find
the bound states of a 2D-SHO. We can then readily check the
validity of our whole procedure, which includes our prescription
for finding the optimal quantities $\hat{L_x}(N)$ and
$\hat{L_y}(N)$, and the overall accuracy of our results.

\begin{figure}
    \centering
  \includegraphics[width=10cm]{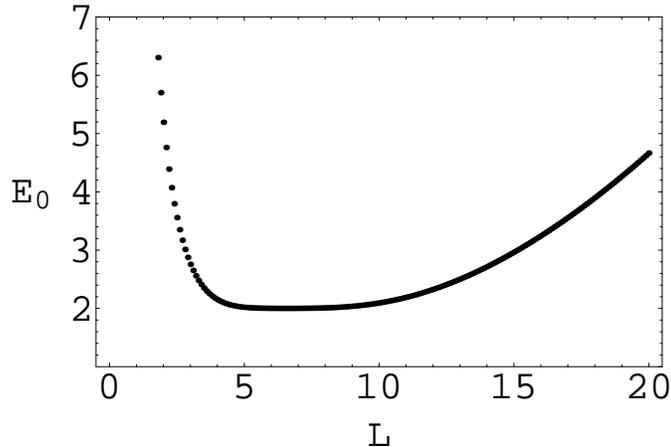}\\
  \caption{Ground state energy  for 2D-SHO versus $L$ for $N=6$, using SM in units where $\hbar\omega=2$. The position of the minimum
  determines $\hat{L}(6)$.}\label{figNL}
\end{figure}

\begin{figure}
  \centering
  \includegraphics[width=10cm]{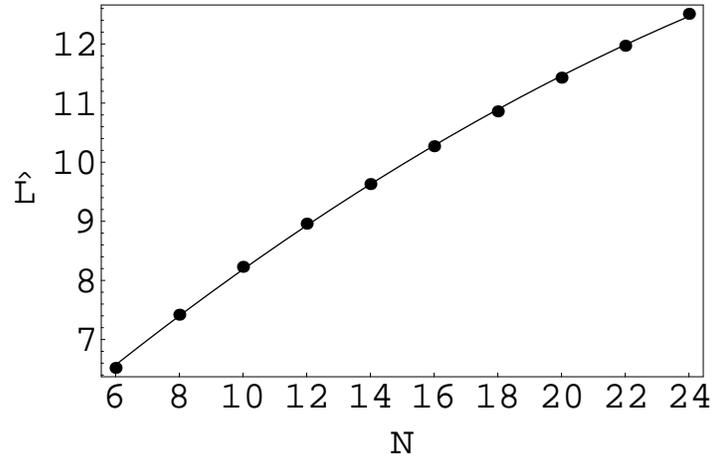}\\
  \caption{The dots represent the values of $\hat{L}$ computed by the method described in the text for different values of $N$.
  The solid line represents the computed
  interpolation function $\hat{L}(N)$.}\label{figLN2}
\end{figure}

\begin{figure}
\centerline{
\includegraphics[width=10cm]{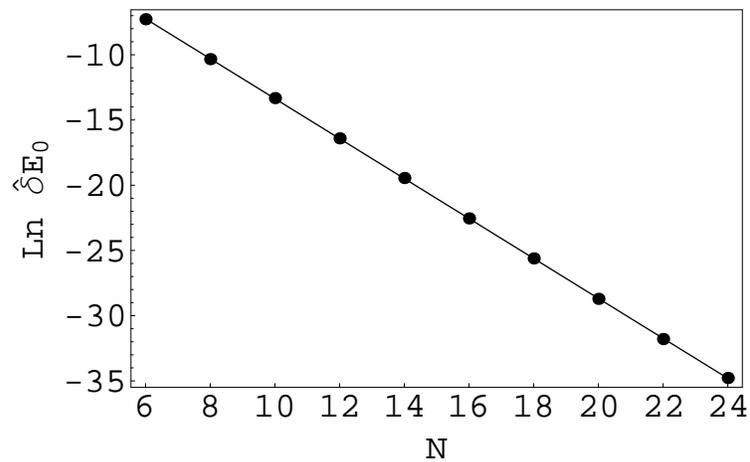}
} \caption{Semi-log plot of the error for the first eigenvalue of
the 2D-SHO obtained by RSM using various number of basis
functions.}\label{fig-error}
\end{figure}

\begin{figure}
\centering
  \includegraphics[width=17cm]{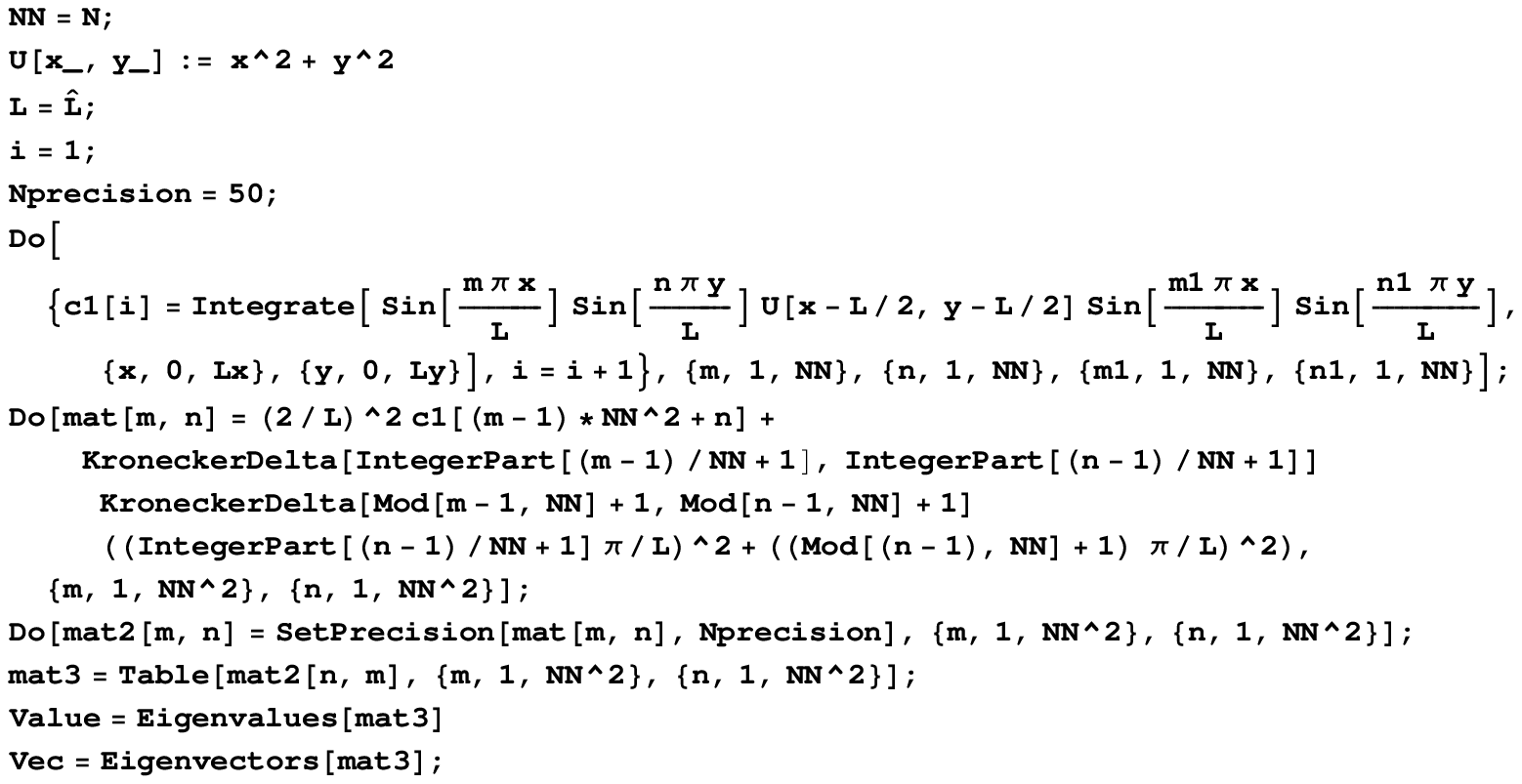}\\
  \caption{MATHEMATICA commands for computing the spectrum of the Hamiltonian $H=p_x^2+p_y^2+U(x,y)$, in
  units where $\hbar=1$. The value for $L$ in line 3 should be obtained by our optimization procedure as described in the text.
  Whenever possible we evaluate the integrals analytically, and replace the Integrate command by its results,
  to increase the precision and save time. This has been the case for the examples presented in this paper.}\label{program}
\end{figure}

The Schr\"{o}dinger equation for an isotropic 2D-SHO is,
\begin{equation}\label{SHO}
-\frac{\hbar^2}{2m}\left(\frac{d^2\psi(x',y')}{dx'^2}+\frac{d^2\psi(x',y')}{dy'^2}\right)+\frac{1}{2}m
\omega^2 (x'^2+y'^2) \psi(x',y')=E'\psi(x',y'),
\end{equation}
where $\omega$ is the natural frequency of the Oscillator. We
first shift the variables as explained above, and then we convert
this differential equation into the following dimensionless form
by dividing both sides by $\hbar\omega/2$,
\begin{equation}\label{SHO2}
-\frac{d^2\psi(x,y)}{dx^2}-\frac{d^2\psi(x,y)}{dy^2}+\left((x-L_x/2)^2+(y-L_y/2)^2\right)
\psi(x,y)=E\psi(x,y),
\end{equation}
where
$x=\sqrt{\frac{m\omega}{\hbar}}x'$,$y=\sqrt{\frac{m\omega}{\hbar}}y'$,
and $E=\frac{2}{\hbar\omega}E'$. This differential equation is
exactly solvable and its eigenvalues and eigenfunctions, which are
all bound states, can be easily found analytically and are well
known,
\begin{eqnarray}
\hspace{0cm}\psi_{n_x,n_y}(x,y)&=&\left(\frac{1}{\pi}\right)^{1/2}\frac{H_{n_x}(
x)H_{n_y}( y)} {\sqrt{2^{n_x+n_y}n_x!n_y!}}e^{-(x^2+y^2)/2},\\
E_{n_x,n_y}&=&2(n_x+n_y+1),\hspace{1cm}n_x,n_y=\{0,1,2,...\},\label{SHO1}
\end{eqnarray}
where $H_n(x)$ denote the Hermite polynomials. Using RSM we can
calculate accurately the energy levels and the corresponding
eigenfunctions of this Hamiltonian. Here, we choose our
optimization procedure for the ground state which will be
symmetric in $x$ and $y$, therefore
$\hat{L}_x(N)=\hat{L}_x(N)\equiv\hat{L}(N)$. The computation of
the errors of the wave functions are analogous to that of the
energy. We divide the configuration space into $M$ grid points.
Then, we average the square of the absolute value of the
difference between the exact solution and that obtained by the RSM
on the grid points,
\begin{equation}
\delta^2_{\psi}=\frac{\sum_{i,j=1}^M
|\psi_{exact}(i,j)-\psi_{N}(i,j)|^2}{\sum_{i,j=1}^M
|\psi_{exact}(i,j)|^2},\hspace{2cm}\delta_E=\frac{|E_n^{exact}-E_n^{SM}|}{E_n^{exact}}.
\label{eqerror}
\end{equation}
In the above equation we have also shown the expression for
$\delta_E$, for ease of reference. Figure \ref{figNL} shows the
ground state energy computed using SM for the fixed value of the
$N=6$ as a function of $L$. Note the existence of the minimum
point at the exact value of the eigenvalue. This point determines
$\hat{L}(6)$. We repeat this procedure for a few other values of
$N$. After plotting these values we can obtain an interpolating
function $\hat{L}(N)$ (Fig. \ref{figLN2}). The optimization method
introduce here is equivalent to the one introduced in Ref.
\cite{RSM}, where inflection points determined the quantities
$\hat{L}(N)$. Table \ref{Table Oscillator} shows the complete
results for the first 10 eigenvalues and eigenvectors for $N=22$.
Several points are note worthy here. First, note the outstanding
accuracy of $\delta_E\approx 10^{-15}$ for the ground state in
particular, and the general good correspondence between $\delta_E$
and $\hat\delta_E$. Also note the corresponding good accuracy for
$\delta_{\psi}$, reported only for the non-degenerate cases. We
did not calculate $\delta_{\psi}$ for other cases because the
outcome of the algorithm in each degenerate subspace causes an
unpredictable linear combination of those wave functions, which is
equivalent to a whole rotation in that subspace. Hence the
computation of $\delta_{\psi}$ becomes a little complicated. Also
note that the errors associated with wave functions symmetric in
$x$ and $y$  are about one order of magnitude better than the
asymmetric ones in the degenerate subspace, because we assumed
this symmetry in our optimization procedure. In Fig.
\ref{fig-error} we show a semi-log plot of the error for the
ground state energy, obtained using RSM, in terms of $N$, all
obtained using appropriate $\hat{L}(N)$. Note that the error falls
off exactly exponentially as a function of $N$, a theoretical
property common to all SM \cite{SP1}. The exact matching of our
computed error with this theoretical expectation is another
positive sign for our method. In Fig. \ref{program} we state the
MATHEMATICA program for solving this problem, to emphasize how
short our program is. We have only left out the the computation of
$\hat{L}(N)$.
\begin{table}
  \centering
\begin{tabular}{|c|c|c|c|c|c|} \hline
  $n_x,n_y$   & $E_n^{exact}$ & $E_n^{SM}$       & $\delta_E$  & $\hat\delta_E$& $\delta_{\psi}$   \\ \hline
   0,0    &2.      &    2.000000000000015572     & $7.79\times10^{-15}$  &  $7.45\times10^{-15}$ & $1.58\times10^{-8}$                 \\ \hline
   0,1    &4.      &    4.000000000000278511     & $6.96\times10^{-14}$  &  $6.68\times10^{-14}$ & -              \\ \hline
   1,0    &4.      &    4.000000000000278512     & $6.96\times10^{-14}$  &  $6.68\times10^{-14}$ &  -      \\ \hline
   1,1    &6.      &    6.000000000000541453     & $9.02\times10^{-14}$  &  $8.63\times10^{-14}$ & $1.90\times10^{-7}$                \\ \hline
   2,0    &6.      &    6.000000000018044778     & $3.00\times10^{-12}$  &  $2.83\times10^{-12}$ &   -          \\ \hline
   0,2    &6.      &    6.000000000018044778     & $3.00\times10^{-12}$  &  $2.83\times10^{-12}$ & -             \\ \hline
   1,2    &8.      &    8.00000000001830772      & $2.29\times10^{-12}$  &  $2.16\times10^{-12}$ & -             \\ \hline
   2,1    &8.      &    8.00000000001830772      & $2.29\times10^{-12}$  &  $2.16\times10^{-12}$ & -             \\ \hline
   3,0    &8.      &    8.00000000019999217      & $2.50\times10^{-11}$  &  $2.38\times10^{-11}$ & -              \\ \hline
   0,3    &8.      &    8.00000000019999217      & $2.50\times10^{-11}$  &  $2.38\times10^{-11}$ & -          \\ \hline
   2,2    &10.     &    10.00000000003607398     & $3.61\times10^{-12}$  &  $3.41\times10^{-12}$ & $8.23\times10^{-7}$             \\ \hline
   1,3    &10.     &    10.00000000020025511     & $2.00\times10^{-11}$  &  $1.90\times10^{-11}$ & -              \\ \hline
   3,1    &10.     &    10.00000000020025511     & $2.00\times10^{-11}$  &  $1.90\times10^{-11}$ & -             \\ \hline
   0,4    &10.     &    10.00000000630282991     & $6.30\times10^{-10}$  &  $5.90\times10^{-10}$ & -               \\ \hline
   4,0    &10.     &    10.00000000630282991     & $6.30\times10^{-10}$  &  $5.90\times10^{-10}$ & -                \\ \hline
   2,3    &12.     &    12.00000000021802137     & $1.81\times10^{-11}$  &  $1.73\times10^{-11}$ & -               \\ \hline
   3,2    &12.     &    12.00000000021802137     & $1.81\times10^{-11}$  &  $1.73\times10^{-11}$ & -            \\ \hline
   1,4    &12.     &    12.00000000630309285     & $5.25\times10^{-10}$  &  $4.24\times10^{-10}$ & -              \\ \hline
   4,1    &12.     &    12.00000000630309285     & $5.25\times10^{-10}$  &  $4.24\times10^{-10}$ & -             \\ \hline
   0,5    &12.     &    12.00000003939548075     & $3.28\times10^{-9}$   &  $3.08\times10^{-9}$  & -            \\ \hline
   5,0    &12.     &    12.00000003939548075     & $3.28\times10^{-9}$    &  $3.08\times10^{-9}$ & -            \\ \hline
   $\hat{L}(22)$&\multicolumn{5}{c|}{$\frac{1197}{100}$}\\ \hline
\end{tabular}
  \caption{The results for the first 21 eigenstates (out of 484) of the 2D-SHO in units where
   $\hbar\omega=2$, using RSM with $N=22$. That is we have used 484 basis functions and $\hat{L}(22)=\frac{1197}{100}$. Note that we obtain all the
   degeneracies and all with very good accuracy. Note that we have good correspondence between  $\delta_E$ and $\hat\delta_E$. The computation time
   was about 697.765 seconds for $N=22$ (using SetPrecision[..., 20] in MATHEMATICA) on a Pentium 2.4 GHz machine resulting in a precision of $10^{-15}$
   for the ground state energy, and it was 11.062 seconds
   for $N=18$ (using default double precision) resulting in a precision of $10^{-12}$. } \label{Table Oscillator}
\end{table}
\pagebreak
\section{An example which is not exactly solvable}
\begin{figure}
\centering \epsfig{figure=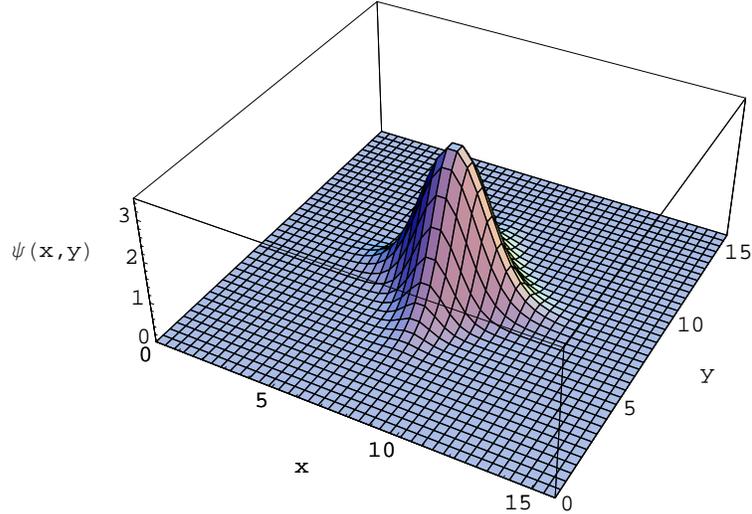,width=10cm}
\caption{The wave function of the ground state of the 2D-QCD
potential using $N=42$ and $\hat{L}(42)=\frac{1553}{100}$, in
units where $\hbar=1$. } \label{figQCD1}
\end{figure}

\begin{figure}
\centerline{\begin{tabular}{ccc} \epsfig{figure=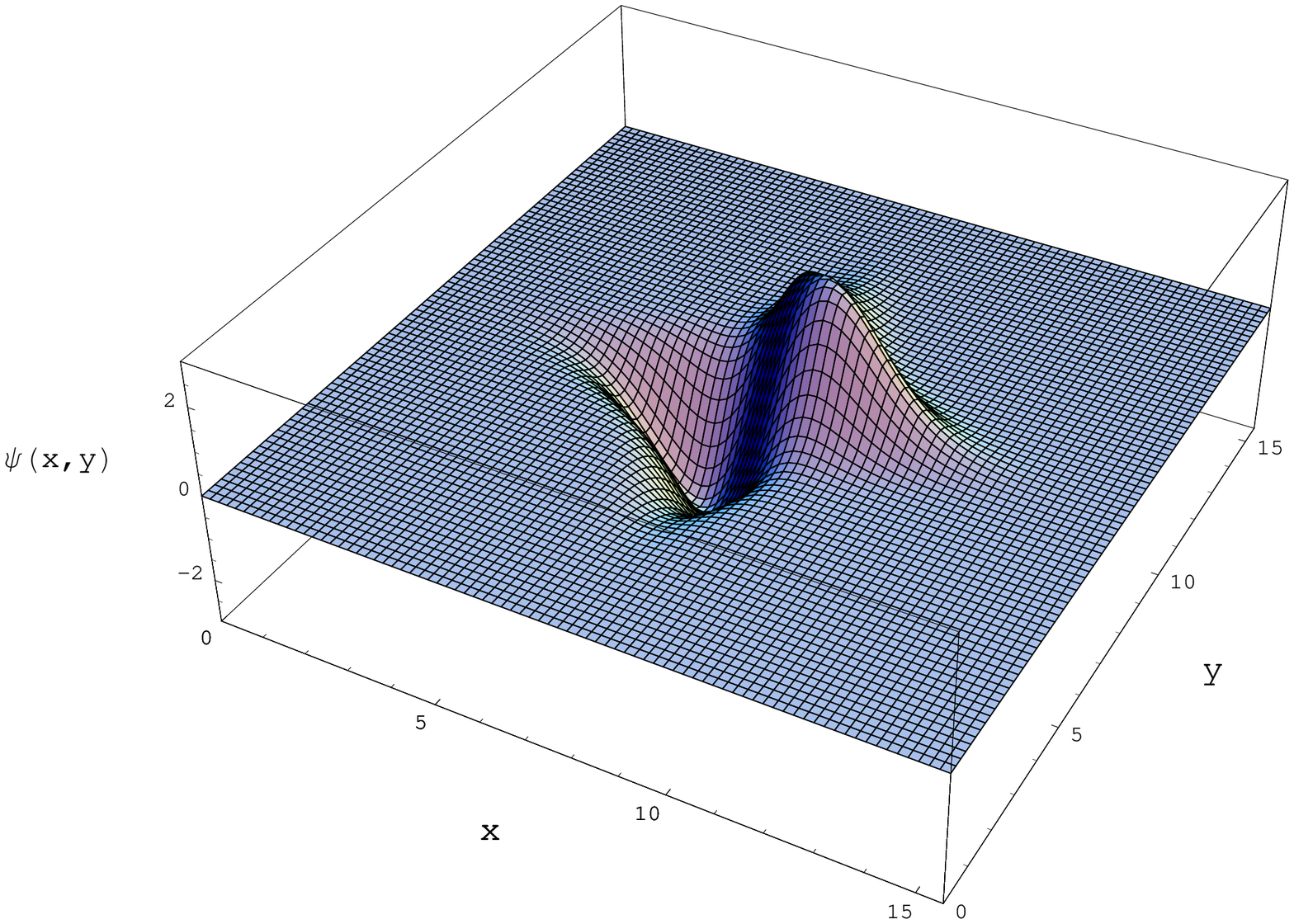,width=8cm}
 &\hspace{2.cm}&
\epsfig{figure=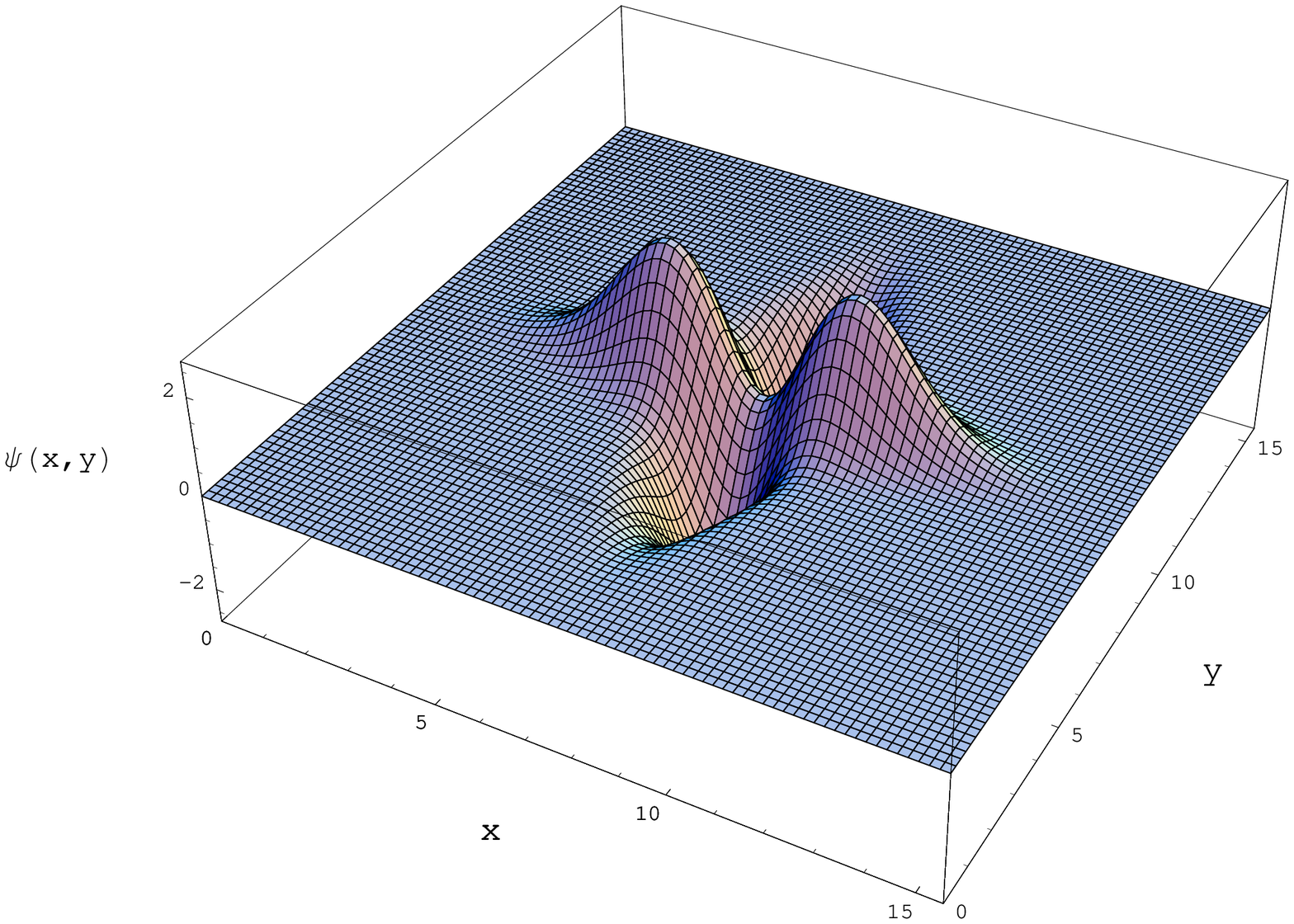,width=8cm}\\
\epsfig{figure=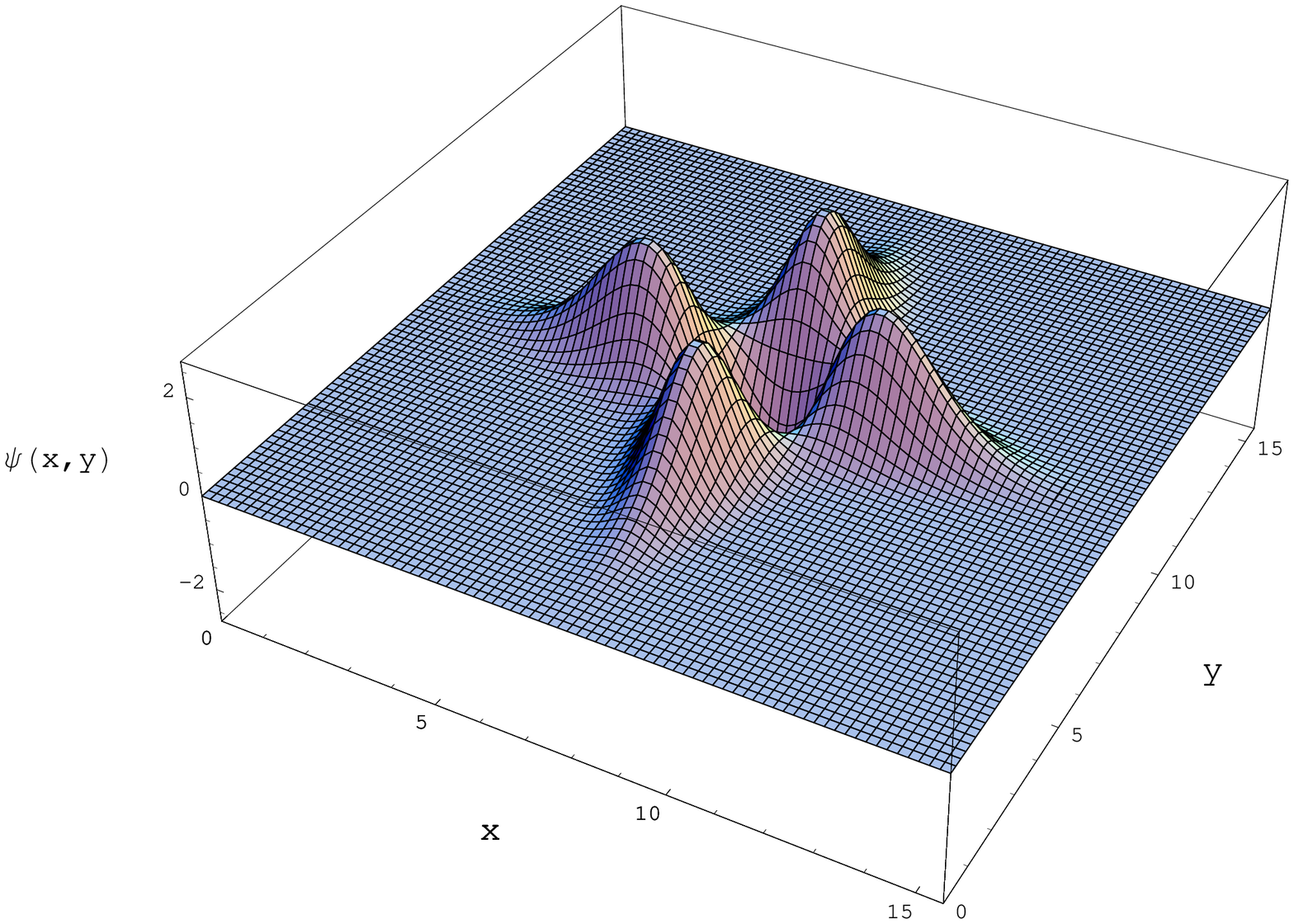,width=8cm}
 &\hspace{2.cm}&
\epsfig{figure=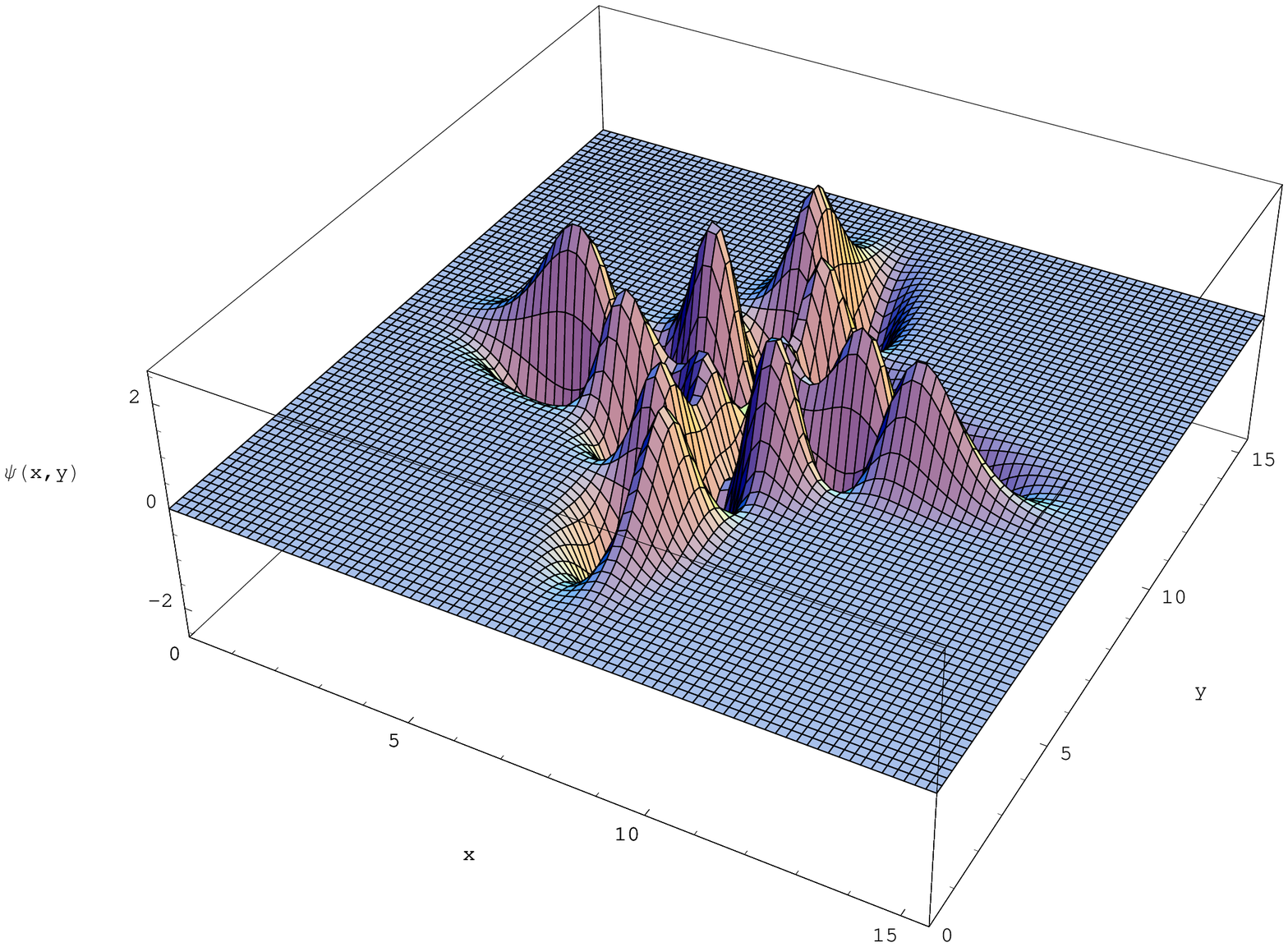,width=8cm}
\end{tabular}}
\caption{Some eigenfunctions of the 2D-QCD potential. Upper left:
2nd; Upper right 4th; Lower left: 5th, and lower right: 44th
states, respectively. All the states are non-degenerate except the
first one. Its degenerate wave function can be obtained by 90
degree rotation. The last one is shown because it is symmetrical,
interesting, and has a relatively low error. Same parameters were
used as for Fig. \ref{figQCD1}} \label{figQCD2}
\end{figure}

The dimensionless and shifted Schr\"{o}dinger equation for the
example that we want to solve here is,
\begin{equation}\label{QCD}
-\left(\frac{d^2\psi(x,y)}{dx^2}+\frac{d^2\psi(x,y)}{dy^2}\right)+\alpha\,\,
(x-L/2)^2 (y-L/2)^2 \psi(x,y)=E\psi(x,y),
\end{equation}
where $\alpha$ is a positive constant. The potential in this
example is sometimes called the 2D-QCD potential. This PDE is
elliptic and not exactly solvable. Therefore, we use RSM to find
its eigenvalues and eigenfunctions. In Table. \ref{TableQCD} we
have shown the eigenvalues for the first 12 states and some other
highly exited ones ($n=\{20,25,33,44\}$). The latter were chosen
for their unusually high accuracy due to their symmetric form, and
the fact that we have chosen
$\hat{L}_x(N)=\hat{L}_y(N)=\hat{L}(N)$.
\begin{table}
  \centering
\begin{tabular}{|c|c|c|} \hline
  $n$    & $E_n^{SM}$       & $\hat\delta_E$   \\ \hline
   1         &   1.10822315780256     & $1.19\times10^{-10}$                     \\ \hline
   2          &    2.37863785124994     & $1.16\times10^{-8}$                    \\ \hline
   3          &    2.37863785124996     & $1.16\times10^{-8}$               \\ \hline
   4          &    3.05608156130323     & $2.06\times10^{-7}$                      \\ \hline
   5         &    3.51495134040797     & $1.12\times10^{-6}$                    \\ \hline
   6          &    4.09348955687600    & $9.38\times10^{-6}$                    \\ \hline
   7          &    4.09348955687604    & $9.38\times10^{-6}$                   \\ \hline
   8         &    4.75298944936096    & $9.32\times10^{-5}$                   \\ \hline
   9          &    4.98538290136962    & $1.75\times10^{-5}$                    \\ \hline
   10          &    5.01127928161308    & $4.59\times10^{-11}$                \\ \hline
   11         &    5.50103621623983     & $7.92\times10^{-4}$                    \\ \hline
   12         &    5.50103621623990     & $7.92\times10^{-4}$                    \\ \hline\hline
   20         &    8.07437393671447     & $6.64\times10^{-9}$                   \\ \hline
   25         &   9.27305945794927     & $3.36\times10^{-8}$                     \\ \hline
   33        &    11.4718771513251     & $7.24\times10^{-7}$                      \\ \hline
   44         &    13.8662683175987    & $8.33\times10^{-6}$                     \\ \hline
   $\hat{L}(42)$&\multicolumn{2}{c|}{$\frac{1553}{100}$}\\ \hline
\end{tabular}
  \caption{The results for the eigenvalues ($E^{SM}_n$) using RSM with $N=42$, for the first 12 states and some other
highly exited and interesting ones (as explained in the text) for
the 2D-QCD
  Hamiltonian, $p_x^2+p_y^2+x^2y^2$, in units where
   $\hbar=1$.} \label{TableQCD}
\end{table}
Figure \ref{figQCD1} shows the ground state wave function. Note
the slight over extension of the wave function in the $x$ and $y$
directions due to the particular form of the potential.  In Fig.
\ref{figQCD2} we show the wave functions for the second, forth,
fifth, and forty forth eigenstates. The third eigenstate is not
shown because it is degenerate with the second and can be obtained
from it by 90 degree rotation. We have decided to show the forty
forth state because we found it interesting and its highly
symmetric form produces an unusually small error, as explained
above.
\section{Conclusions}
We have extended the Refined Spectral Method to two dimensions and
used it as an extremely accurate method for obtaining the energies
and wave functions of the bound states of the two dimensional
time-independent Schr\"{o}dinger equation. In this method a finite
basis is used for approximating the solutions. The refinement of
the method is accomplished by calculating an optimized spatial
domain for a given number of basis elements, denoted by
$\hat{L}(N)$. The criteria for this optimization is to minimize
the energy for one of the eigenstates, usually chosen to be the
ground state. Note that our refined method is not quite equivalent
to the Rayleigh-Ritz variational method, in that  we have combined
the spectral method in which the wave function is expanded in an
arbitrary basis with optimization of the spatial domain which is
equivalent to adjusting the effective potential for the bound
states. This effective potential for the case of the Fourier basis
with confinement boundary condition is the actual potential plus
the confining ``walls'' placed at the boundaries, which are
separated by $\hat{L}(N)$. This refinement scheme usually improves
the accuracy of SM drastically and this effect increases rapidly
with $N$. We applied this method to an exactly solvable problem
and easily found an extraordinarily good agreement with the exact
solutions (errors of order $10^{-15}$ with only 22 basis
functions). This method is very simple, fast, extremely accurate
in most cases, very robust, stable, and there is no need to
specify the boundary conditions on the slopes. Most importantly,
one can obtain the energies and the wave functions of as many of
the bound states as desired with a single run of the algorithm.
The generalization of this method to higher dimensional cases is
straight forward.
\vskip20pt\noindent {\large {\bf
Acknowledgement}}\vskip5pt\noindent The Authors thank A. Turbiner
for his useful suggestion regarding the QCD potential. This
research has been supported by the office of research of Shahid
Beheshti University under Grant No. 500/3787. \vskip10pt


\begin{thebibliography}{99}
\bibitem{Schrodinger}E. Schr\"{o}dinger, Quantisierung als Eigenwertproblem. (Erste Mitteilung.),
Ann. Phys. (Leipzig) 79 (1926), 361–376; Quantisierung als Eigenwertproblem. (Zweite Mitteilung.),
Ann. Phys. (Leipzig) 79 (1926), 489–527; Quantisierung als Eigenwertproblem. (Dritte Mitteilung.),
Ann. Phys. (Leipzig) 80 (1926), 437–490; \"{U}ber das Verh\"{a}ltnis der Heisenberg-Born-
Jordan'schen Quantenmechanik zu der meinen. Ann. Phys. (Leipzig), 79 (1926),
734-756; An Undulatory Theory of the Mechanics of Atoms and Molecules, Phys. Rev.
28, 1049 (1926).
\bibitem{Infeld}L. Infeld and T.D. Hull, The Factorization Method, Rev. Mod. Phys. 23, 21 (1951).
\bibitem{factorization}E. Schr\"{o}dinger, A method of determining quantum mechanical eigenvalues and eigenfunctions,
Proc. R. Ir. Acad. Sect. A, Math. Astron. Phys. Sci. 46, 9–16 (1940);
Further studies on solving eigenvalue problems byfactorization, 47A, 183–206 (1941).
\bibitem{Darboux}G. Darboux, Sur une proposition relative aux \'{e}quations lin\'{e}arires, C R. Acad. Sci. III 94, 1456 (1882).
\bibitem{Runge-Kutta}G. Avdelas, T.E. Simos, and J. VigoAguiar, An embedded exponentially-fitted Runge-Kutta method for the numerical solution of
the Schr\"{o}dinger equation and related periodic initial-value,
problems Comput. Phys. Commun. 131, 52 (2000).
\bibitem{Numerov}G. Avdelas and T.E. Simos, Dissipative high phase-lag order Numerov-type methods for the numerical solution of the
Schr\"{o}dinger equation, Phys. Rev. E 62, 1375 (2000).
\bibitem{Relax}J.D. Praeger, Relaxational approach to solving the Schr\"{o}dinger
equation, Phys. Rev. A 63, 022115 (2001).
\bibitem{Henyey}L.G. Henyey, L. Wilets, K.H. B\"{o}hm, R. Lelevier, and R.D.
Lev\'{e}e, A method for automatic computation of stellar
evolution, Astrophys. J. 129, 628 (1959).
\bibitem{hierarchical}M. Sugawara, Adaptive basis set for quantum mechanical calculation based on
hierarchical finite element method,
Chem. Phys. Lett. 295, 423 (1998).
\bibitem{microgenetic}H. Nakanishi and M. Sugawara, Numerical solution of the Schr\"{o}dinger equation by a microgenetic algorithm,
Chem. Phys. Lett. 327, 429 (2000).
\bibitem{Holland}J. H. Holland, Adaptation in Natural and Artificial Systems
(University of Michigan Press, Ann Arbor, 1975, 1992).
\bibitem{RSM}P. Pedram, M. Mirzaei, and S. S. Gousheh, math-ph/0611008
\bibitem{SP1}J. P. Boyd,  Chebyshev \& Fourier Spectral Methods, DOVER Publications, Inc. (2000).
\bibitem{RSM2}P. Pedram, M. Mirzaei, and S. S. Gousheh, math-ph/0611033
\end{thebibliography}
\end{document}